\documentclass[amsmath,amssymb,prc,final,showpacs,twocolumn]{revtex4} 

\usepackage{bm,hyphenat,xspace} 
\usepackage{graphicx,epsfig}

\newcommand{\scl}{0.63}

\newcommand{\C}{{}^{14}\mathrm{C}}
\newcommand{\Cn}{{}^{15}\mathrm{C}}
\newcommand{\Be}{{}^{10}\mathrm{Be}}
\newcommand{\Ben}{{}^{11}\mathrm{Be}}
\newcommand{\A}[2]{{}^{#1}\mathrm{#2}}

\newcommand {\mbf}[1]{{\mathbf{#1}}}
\newcommand {\vecg}[1]{\mbox{\boldmath{$#1$}} }

\begin{document}

\title {
Deuteron stripping and pickup involving halo nuclei $\Ben$ and $\Cn$}

\author{A.~Deltuva} 
\affiliation{Centro de F\'{\i}sica Nuclear da Universidade de Lisboa, 
P-1649-003 Lisboa, Portugal }

\received{11 March 2009}


\begin{abstract}
Three-body calculations of $(d,p)$ and $(p,d)$ reactions involving 
one-neutron halo nuclei $\Ben$ and $\Cn$ are performed 
using the framework of Faddeev-type scattering equations.
Important effects of the optical potential nonlocality are found
improving the description of the experimental data.
The obtained values for the neutron spectroscopic factor
are consistent with estimations from other approaches.
\end{abstract}

 \maketitle

\section{Introduction \label{sec:intro}}

Deuteron stripping and pickup reactions $(d,p)$ and $(p,d)$
constitute an important tool for 
extracting  nuclear structure information such as 
spectroscopic factors or spin/parity.
Of special interest are cases involving exotic weakly bound systems
such as one-neutron halo nuclei assumed to consist of a core 
with a mass number $A$ and a 
neutron $n$; well known examples are $\Ben$, $\Cn$, and $\A{19}{C}$.
 The reaction $d+A \to p+(An)$ and its time reverse $p+(An) \to d+A $
may therefore be described in a three-body model $(p,n,A)$.
These reactions have been analyzed by several approximate
methods such as distorted-wave Born approximation (DWBA) 
and various DWBA-type adiabatic approaches 
\cite{johnson:70a,timofeyuk:99a,liu:04a,lee:07a,pang:07a} 
or continuum-discretized coupled-channels (CDCC) method 
\cite{austern:87,moro:02a,ogata:03a,moro:06a}.
Recently, also the application of exact three-body Faddeev/Alt, 
Grassberger, and Sandhas (AGS) scattering theory \cite{faddeev:60a,alt:67a}
 to three-body nuclear reactions has become possible
\cite{deltuva:06b,deltuva:07d} 
due to a novel implementation \cite{deltuva:05c} of the screening and
renormalization method \cite{taylor:74a,alt:80a,deltuva:08c}
for including the long-range Coulomb force between charged particles.
Faddeev/AGS framework, though being technically and computationally
most complicated and expensive, has an advantage that, once numerically 
well-converged 
results are obtained, all discrepancies with the experimental data can be 
attributed solely to the shortcomings of the used optical potentials (OP) 
or to the inadequacy of the three-body model.
Recent comparison \cite{deltuva:07d} of Faddeev/AGS and CDCC results
revealed that CDCC is indeed a reliable method to calculate $d+A$ 
elastic and  breakup cross sections but may lack
accuracy  for transfer reactions such as $p+ \Ben \to d+ \Be$.
No benchmark with adiabatic approaches  is performed yet.
Furthermore, the nonlocal optical potential (NLOP), so far, has been included
only in the Faddeev/AGS
 framework \cite{deltuva:09b} where an important effect of the OP
nonlocality was found in deuteron stripping on $\A{12}{C}$ and $\A{16}{O}$,
improving the description of the experimental data.
Therefore we expect the OP nonlocality to be significant also in transfer
reactions on one-neutron halo nuclei which we aim to study
in the present work using momentum-space AGS equations.

The theoretical framework is shortly recalled in Sec.~\ref{sec:th}.
The results for $(d,p)$ and $(p,d)$ reactions involving $\Ben$ and $\Cn$ nuclei
for which the experimental data is available
are presented in Sec.~\ref{sec:res}. The summary is given
in Sec.~\ref{sec:sum}.

\section{AGS equations} \label{sec:th}

The AGS equations \cite{alt:67a} 
\begin{equation}  \label{eq:Uba}
U_{\beta \alpha}(Z)  = \bar{\delta}_{\beta\alpha} \, G^{-1}_{0}(Z)  +
\sum_{\gamma}   \bar{\delta}_{\beta \gamma} \, T_{\gamma}(Z) 
\, G_{0}(Z) U_{\gamma \alpha}(Z),
\end{equation}
are a system of Faddeev-like coupled integral equations 
for the transition operators $U_{\beta \alpha}(Z)$; their on-shell matrix 
elements $\langle\psi_{\beta}|U_{\beta \alpha}(E+i0)|\psi_{\alpha}\rangle$
at the available three-particle energy $E$ 
are amplitudes for all scattering  processes (elastic, inelastic, transfer,
breakup) allowed by the chosen Hamiltonian $H = H_0 + \sum_{\gamma} v_{\gamma}$
where $H_0$ is the free Hamiltonian and 
$v_{\gamma}$ the potential for the pair $\gamma$ in odd-man-out notation.
In Eq.~(\ref{eq:Uba}) $ \bar{\delta}_{\beta\alpha} = 1 - \delta_{\beta\alpha}$,
$G_0(Z) = (Z-H_0)^{-1}$ is the free resolvent, and 
\begin{equation}  \label{eq:T}
T_{\gamma}(Z) = v_{\gamma} + v_{\gamma} G_0(Z) T_{\gamma}(Z)
\end{equation}
is the two-particle transition matrix.
The channel states $|\psi_{\gamma}\rangle$ for  $\gamma = 1,2,3$ are the
eigenstates of the corresponding channel Hamiltonian $H_\gamma = H_0 + v_\gamma$
with the energy eigenvalue $E$; thus, $|\psi_{\gamma}\rangle$ is a product of
the bound state wave function for pair $\gamma$ and a plane wave
with fixed on-shell momentum $\mbf{q}_\gamma$
corresponding to the relative motion of particle $\gamma$ and pair $\gamma$
in the initial or final state. 

The AGS equations (\ref{eq:Uba}) are applicable only to short-range 
interactions.
Nevertheless, the long-range Coulomb force between charged particles can
be included in this framework using the method of screening and renormalization
\cite{taylor:74a,alt:80a,deltuva:08c} where one has to solve AGS equations 
with nuclear plus screened Coulomb potential to obtain
the Coulomb-distorted short-range part of the transition amplitude;
the convergence of the results with the screening radius has to be established.
The method has been successfully applied to proton-deuteron elastic scattering 
and breakup \cite{deltuva:05c,deltuva:05a}
and to direct nuclear reactions dominated by three-body degrees of freedom
 \cite{deltuva:06b,deltuva:07d}.

We use momentum-space partial-wave basis.
The screened Coulomb potential and most of the nuclear potentials are 
given in coordinate space; they have to be transformed to the momentum space 
where local and nonlocal potentials are treated in the same manner.
The  AGS equations then become a system of integral equations
with two continuous variables which are the absolute values of Jacobi momenta.
The technique of numerical solution is described in 
Refs.~\cite{chmielewski:03a,deltuva:03a,deltuva:05a}.

\section{Results} \label{sec:res}

The dynamic input to AGS equations are the potentials $v_{\gamma}$
for the three pairs of particles.
For the $np$ interaction we take the realistic
CD Bonn potential \cite{machleidt:01a}.
In order to describe reactions involving one-neutron halo nucleus
we need a real $nA$ potential that reproduces the bound state spectrum
of nucleus $(An)$. This is a standard choice in $p+(An)$ reactions 
where the $pA$ potential is complex and is taken at the proton lab energy.
As discussed in Ref.~\cite{deltuva:09b}, $d+A \to p+(An)$ and $p+(An) \to d+A$
reactions are related by time reversal provided the energy in the center of mass
(c.m.) system is the same. Therefore we can calculate the latter one using
the standard Hamiltonian with the
nucleus $(An)$ being in its ground or excited state and apply the detailed
balance to obtain the observables for the former one.
This is equivalent to calculating the $d+A \to p+(An)$ transfer with a real 
$nA$ potential and complex $pA$ interaction which is a nonstandard choice.
Nevertheless, it provides quite a reasonable description of the
$d+A$ elastic scattering, as demonstrated in Ref.~\cite{deltuva:09b}
and in the present work.
Therefore we use a real partial-wave dependent
$nA$ potential that has local central and spin-orbit parts,
\begin{equation}  \label{eq:VnA}
v_{\gamma}(r) = -V_c f(r,R,a) + \vecg{\sigma}\cdot \mbf{L} \,
V_{so} \, \frac{2}{r} \frac{d}{dr}f(r,R,a),
\end{equation}
with $f(r,R,a) = [1+\exp((r-R)/a)]^{-1}$ and $R = r_0 A^{1/3}$.
The $n$-$\Be$ potential is taken from Ref.~\cite{cravo:09a} while
the one for $n\text{-}\C$ is constructed in the present work assuming standard
values for  $r_0 = 1.25$ fm, $a=0.65$ fm, and  $V_{so}=5.5$ MeV
and adjusting $V_c$ to the binding energy of $\Cn$ ground and excited state
and to the neutron separation energy of  $\C$.
For completeness the parameters of both potentials are given in 
Table~\ref{tab:V} while the resulting values for binding energies 
are listed in Table~\ref{tab:EB}; Pauli forbidden bound states are 
projected out as described in Ref.~\cite{deltuva:06b}.

\begin{table} [b]
\caption{\label{tab:V} Partial-wave dependent strengths of central and 
spin-orbit parts
of $n\text{-}\Be$ and  $n\text{-}\C$ potentials, all in units of MeV.
Other parameters: $r_0 = 1.25$ fm, while $a=0.67$ and 0.65 fm for
$n\text{-}\Be$ and  $n\text{-}\C$ , respectively.}
\begin{ruledtabular}
\begin{tabular}{l*{4}{c}}
$L$ & $V_c(n\text{-}\Be)$ &  $V_{so}(n\text{-}\Be)$ & $V_c(n\text{-}\C)$ &
$V_{so}(n\text{-}\C)$ \\ \hline
0  & 56.413 &   & 50.29  &   \\
odd & 42.498 & 11.953 & 46.13 & 5.50    \\
even, $\geq 2$  & 56.413 & 5.38 & 49.14  & 5.50    \\
\end{tabular}
\end{ruledtabular}
\end{table}

\begin{table} [!]
\caption{\label{tab:EB} Binding energies (MeV) of the bound states corresponding
to the potential parameters of Table~\ref{tab:V}.
Pauli forbidden bound states that are removed are marked with *.}
\begin{ruledtabular}
\begin{tabular}{l*{3}{r}*{2}{l}}
 & $1s_{1/2}$  & $2s_{1/2}$  & $1p_{3/2}$  & $1p_{1/2}$  & $1d_{5/2}$ \\ \hline
$\A{11}{Be}$ & 28.730* & 0.503 & 6.812* & 0.183 &  \\
$\A{15}{C}$ & 28.194* & 1.218 & 11.522* & 8.175* & 0.479 \\
\end{tabular}
\end{ruledtabular}
\end{table}

For the hadronic part of the $pA$ interaction we take the
NLOP of Giannini et al. \cite{giannini2} which is based on Watson's
multiple scattering theory; the nonlocality arises mainly due to the
fully off-shell two-nucleon transition matrix \cite{giannini1,elster:89a}.
In the configuration space the NLOP is parametrized as
\begin{gather}  \label{eq:V}
\begin{split}
v_{\gamma}(\mbf{r}',\mbf{r}) = {} & H_c(x)U_c(y) + H_t(x)U_t(y)\frac{A-2Z}{A} \\ 
& +  H_s(x) V_s(y) \vecg{\sigma}\cdot \mbf{L} 
\end{split}
\end{gather} 
with $x = |\mbf{r}'-\mbf{r}|$, $y=|\mbf{r}'+\mbf{r}|/2$, 
$H_j(x) = (\pi \beta_j^2)^{-3/2} \exp{(-x^2/\beta_j^2)}$,
and $Z$ being the number of protons in the nucleus $A$.
The values of the nonlocality parameters are
$\beta_c = 1.015$ fm, $\beta_t = 1.633$ fm, $\beta_s = 0.789$ fm.
The nonlocality of the OP mainly
Both $U_c(y)$ and  $U_t(y)$ contain real volume and imaginary surface parts,
while the spin-orbit part $V_s(y)$ is real; all of them are parametrized in the 
standard way using Woods-Saxon functions and their derivatives~\cite{giannini2},
i.e.,
\begin{subequations}
\begin{align}
U_j(y) = {} & -V_j f(y,a_R) -4iW_jf(y,a_I)[1-f(y,a_I)], \\
 V_s(y)= {} & V_S \frac{2}{y} \frac{d f(y,a_R)}{dy}, \\
f(y,a_k) = {} & [1+\exp{((y-R_N)/a_k)}]^{-1}
\end{align}
\end{subequations}
with $R_N=1.16A^{1/3}$ fm, $a_R = 0.57$ fm, $a_I = 0.54+0.0032A$ fm,
$V_c = 85$ MeV, $V_t = 127+11ZA^{-1/3}$ MeV, $V_S = 9.1$ MeV,
$W_t = 13$ MeV, and $W_c = w_N[1-\exp(-0.05E)]$ MeV where
$E= E_p^{\mathrm{c.m.}} +1.08 - 1.35ZA^{-1/3}$ and $E_p^{\mathrm{c.m.}}$ is
the proton energy (MeV) in the c.m. frame. We adjust  $w_N$
 to improve the description of the experimental $pA$ scattering data 
in the energy regime relevant for the considered three-body reactions.
In the case of $p$-$\C$ we fit the NLOP 
to the data at proton lab energy $E_p = 14.5$ MeV \cite{p14C14}.
We need the $p$-$\Be$ potential in a broader energy range than
$E_p = 12 - 16$ MeV where experimental data \cite{dBe12p} is available.
Since  the local energy-dependent OP by Watson {\it et al.}~\cite{watson}
provides rather satisfactory description of that data
as demonstrated in Ref.~\cite{dBe12p} and confirmed by our own calculations,
 we adjust the NLOP to the predictions of Watson OP.
 The obtained values for $w_N$ are  given in Table \ref{tab:W};
other parameters are taken from Ref.~\cite{giannini2}.
This time we do not use the local OP obtained from NLOP by equivalence 
transformation \cite{giannini2}; instead, the Watson OP that is often used
also in DWBA-type and CDCC calculations serves as a local reference potential.

Although some parameters of those $pA$ potentials vary with energy, in
the calculations of present paper based on
the Hamiltonian theory for three-body scattering they are taken at a fixed 
energy, corresponding to the proton lab energy in the $p+(An)$ reaction. 
Therefore the chosen Hamiltonian with complex $pA$ potential prevents the 
calculation of $d+A \to n+(Ap)$ and $p+(An) \to n+(Ap)$ reactions since the 
nucleus $(Ap)$ is not bound. These reactions can be described \cite{deltuva:09a}
allowing for an explicitly energy-dependent $pA$ potential which becomes 
real at negative relative  $pA$ energies and reproduces  bound states of the
nucleus $(Ap)$. However, such an approach is not free of theoretical
complications due to the absence of the Hamiltonian theory as discussed
in Ref.~\cite{deltuva:09a}; we therefore refrain from using it in the present
work. Furthermore, the results of Ref.~\cite{deltuva:09a} indicate that
the effect of the energy-dependence in the $pA$ potential 
on the $(d,p)$ and $(p,d)$ cross section
is rather insignificant at small c.m. scattering angles up to 
$\Theta_{c.m.} \approx 40^{\circ}$, although it may become sizable
for large angles where the cross section itself is small.

\begin{table} [!]
\caption{\label{tab:W} 
The parameter $w_N$ of NLOP adjusted to the $p$-$\C$ data 
and  $p$-$\Be$ Watson OP predictions at given proton lab energies $E_p$
(all in units of MeV).} 
\begin{ruledtabular}
\begin{tabular}{l*{2}{r}}
& $E_p$ & $w_{N}$ \\ \hline
$p$-$\Be$ & 9.0 & 40.0     \\
$p$-$\Be$ & 12.0 & 36.0     \\
$p$-$\Be$ & 20.9 & 27.0     \\
$p$-$\Be$ & 35.3 & 18.0     \\
$p$-$\C$  & 14.0 & 28.0     \\
\end{tabular}
\end{ruledtabular}
\end{table}

The interaction between $np$, $nA$, and $pA$ pairs is included in partial
waves with pair orbital angular momentum $L \leq 3$, 6, and 12, respectively,
and the total angular momentum is $J \leq 25$; the results are well converged.
The $pA$ channel is more demanding than the $nA$ channel due to the screened
Coulomb force, where the screening radius $R \approx 10$ to 12 fm for the
short-range part of the scattering amplitude is sufficient for convergence.

\renewcommand{\scl}{0.56}
\begin{figure}[!]
\begin{center}
\includegraphics[scale=\scl]{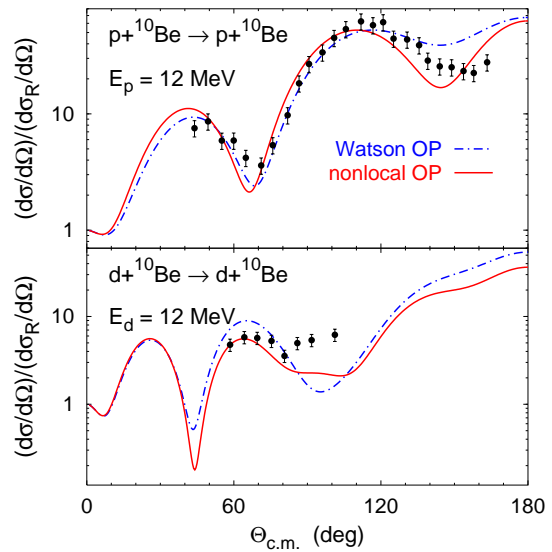}
\end{center}
\caption{\label{fig:pdBe}  (Color online)
Differential cross section divided by Rutherford cross section
for  $p + \Be$ elastic scattering at $E_p = 12$ MeV
and  $d + \Be$ elastic scattering at $E_d = 12$ MeV.
Predictions of NLOP (solid curve) and Watson OP (dashed-dotted curve)
are compared with the experimental data from Ref.~\cite{dBe12p}.}
\end{figure}

\begin{figure}[!]
\begin{center}
\includegraphics[scale=\scl]{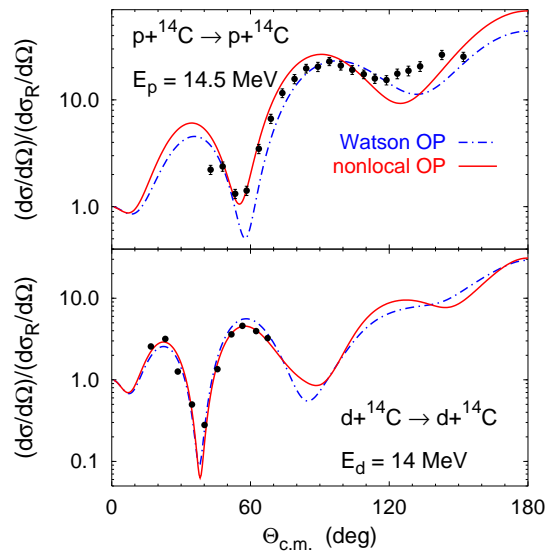}
\end{center}
\caption{\label{fig:pdC}  (Color online)
Differential cross section divided by Rutherford cross section
for  $p + \C$ elastic scattering at $E_p = 14.5$ MeV
and  $d + \C$ elastic scattering at $E_d = 14$ MeV.
Curves as in  Fig.~\ref{fig:pdBe} and data from Refs.~\cite{p14C14,d14C14p}.}
\end{figure}

Examples for the cross section in two-body $p+\Be$ and $p+\C$ 
 and three-body $d+\Be$ and $d+\C$  elastic
scattering are presented in Figs.~\ref{fig:pdBe} and \ref{fig:pdC}
and compared with the available experimental data.
Two-body results indicate that adjusting only one parameter $w_N$ of NLOP 
is sufficient to achieve the quality  in fitting the $pA$ data 
comparable with the one of the Watson OP. However, the three-body $d+A$ data
seems to be reproduced better by the NLOP, although it also fails 
for $d + \Be$ at large angles. There is no $p+(An)$ elastic scattering
data in the considered energy region; however, as demonstrated in 
Ref.~\cite{deltuva:09b}, the theoretical predictions for $p+A$ and $p+(An)$ 
elastic scattering observables are strongly correlated and therefore
also the nonlocality effect on $p+(An)$ elastic cross section is extremely 
small. Thus, the observed discrepancy in $p+\Ben$ elastic cross section
at higher energy \cite{deltuva:07d,lapoux:Be} cannot be resolved by the NLOP.

\renewcommand{\scl}{0.6}
\begin{figure}[!]
\begin{center}
\includegraphics[scale=\scl]{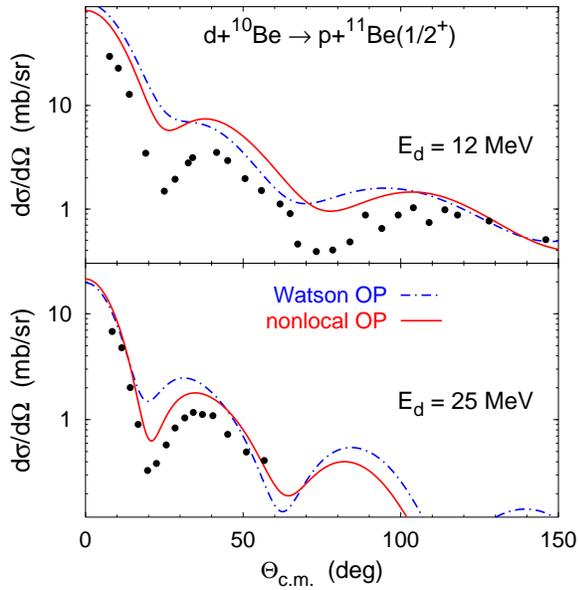}
\end{center}
\caption{\label{fig:dBeps}  (Color online)
Differential cross section for  $d + \Be \to p+\Ben$ transfer to the
$\Ben$ ground state $1/2^+$ at $E_d = 12$ and 25 MeV
as function of the c.m. scattering angle.
Curves as in  Fig.~\ref{fig:pdBe} and the data are from 
Refs.~\cite{dBe12p,dBe25p}.}
\end{figure}

\begin{figure}[!]
\begin{center}
\includegraphics[scale=\scl]{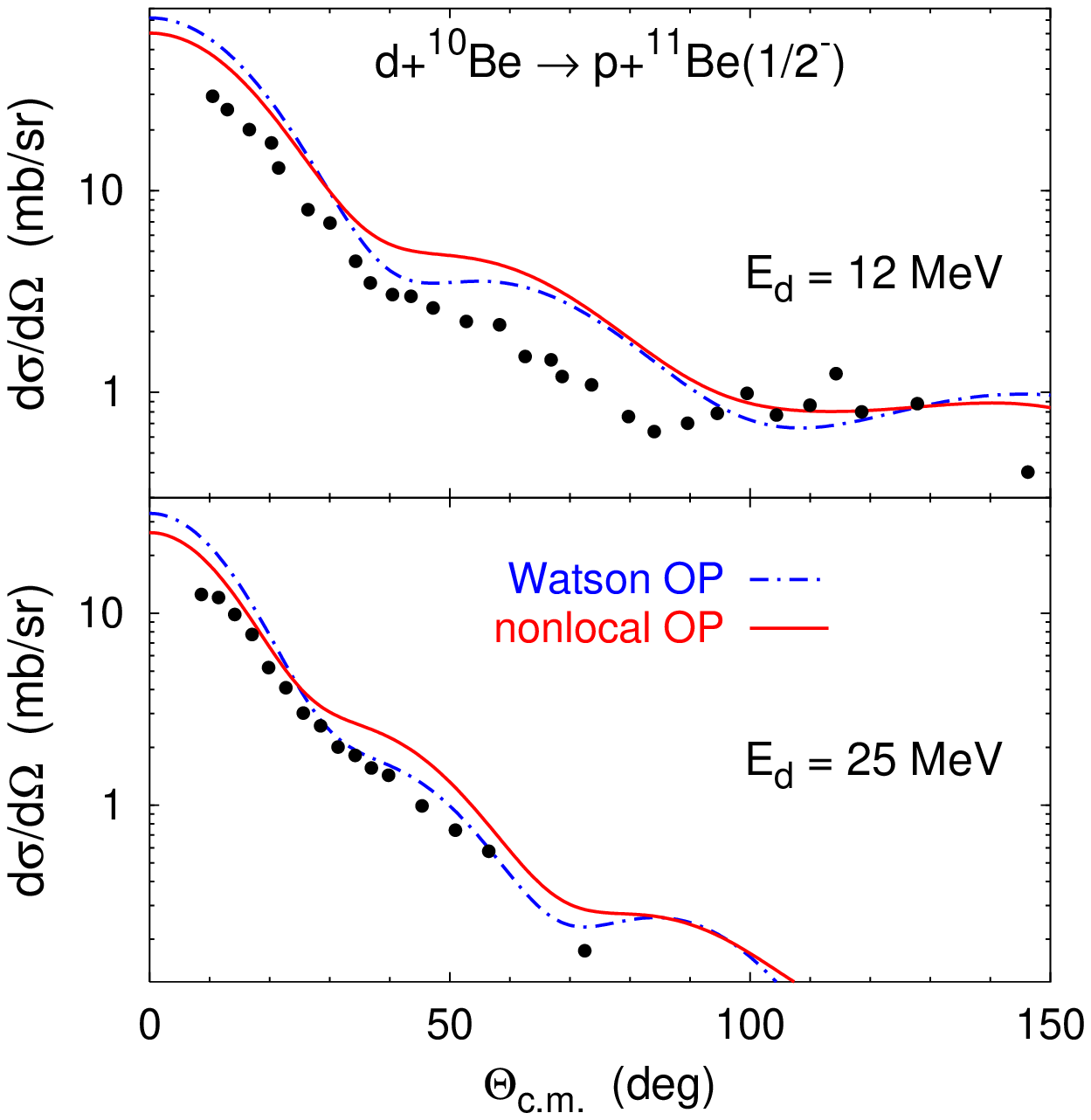}
\end{center}
\caption{\label{fig:dBepp}  (Color online)
Same as  Fig.~\ref{fig:dBeps}, but for  transfer to the $\Ben$ 
excited state $1/2^-$.}
\end{figure}

Results for $d+\Be \to p+\Ben$ transfer at deuteron lab energy 
$E_d= 12$ and 25 MeV are presented in Fig.~\ref{fig:dBeps} for the $\Ben$ 
ground state $1/2^+$ and in Fig.~\ref{fig:dBepp} for the excited state $1/2^-$.
The corresponding proton lab energy in the time-reverse  reaction
$p+\Ben \to d+\Be$ is $E_p = 9.0$ (8.7) and 20.9 (20.5) MeV
for the ground (excited) state, respectively.
At least for the  transfer to the ground state $1/2^+$  the 
 NLOP reproduces the shape of the data clearly better than Watson OP.
In all cases theory overpredicts the data and thereby yields neutron 
spectroscopic factors below 1. This is believed to be due to neglecting
the $\Be$ core excitation in the $\Ben$ bound state \cite{winfield:01}. 
For the ground state $1/2^+$, the spectroscopic factor value around 0.45 at 
12 MeV is in agreement with the DWBA-type result of Ref.~\cite{lee:07a},
while at 25 MeV the value around 0.75 is consistent with most other 
estimations collected in Ref.~\cite{winfield:01}. 
Slightly larger values of the neutron 
spectroscopic factor are obtained for the excited state $1/2^-$,
around 0.6 and 0.85 at 12 and 25 MeV, respectively.
In  Fig.~\ref{fig:pBe35d} results for $p+\Ben \to d+\Be$ transfer at 
$E_p= 35.3$ MeV  are presented. Again significant difference between
NLOP and Watson OP predictions is found. The values for the $\Ben$ 
ground state neutron spectroscopic factor obtained using the NLOP from
$(d,p)$ reaction at $E_d= 25$ MeV and $(p,d)$ reaction at $E_p= 35.3$ MeV 
are consistent with each other, in contrast to those obtained using
the Watson OP.

\renewcommand{\scl}{0.64}
\begin{figure}[t]
\begin{center}
\includegraphics[scale=\scl]{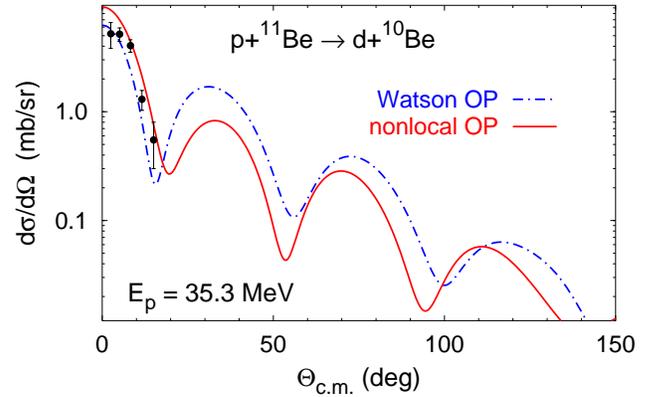} 
\end{center}
\caption{\label{fig:pBe35d} (Color online)
Differential cross section for $p+\Ben \to d+\Be$ transfer at $E_p=35.3$ MeV.
Curves as in Fig.~\ref{fig:pdBe} and the data are from 
Ref.~\cite{winfield:01}.}
\end{figure}

\renewcommand{\scl}{0.75}
\begin{figure*}[!]
\begin{center}
\includegraphics[scale=\scl]{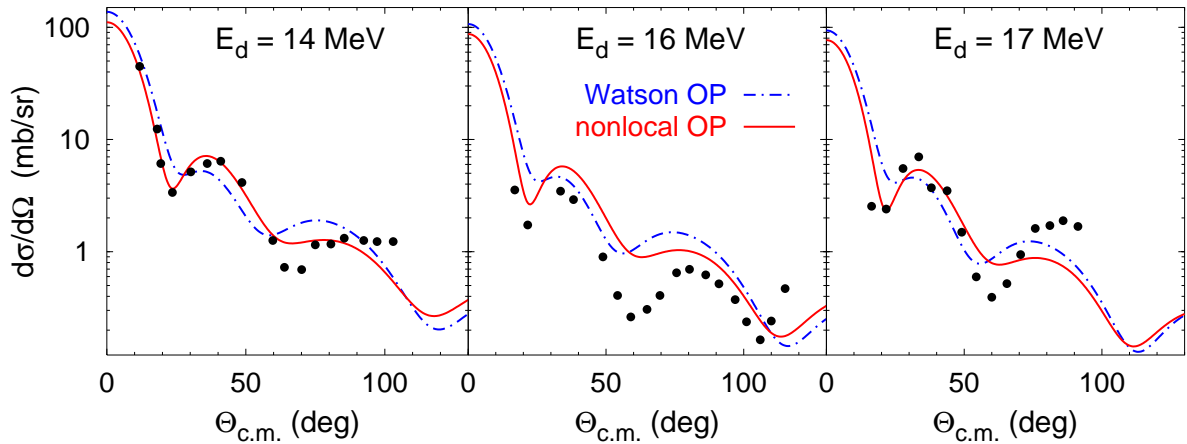}
\end{center}
\caption{\label{fig:dCps}  (Color online)
Differential cross section for  $d + \C \to p+\Cn$ transfer to the
 $\Cn$ ground state $1/2^+$ at $E_d = 14$, 16, and 17 MeV.
Curves as in Fig.~\ref{fig:pdBe} and  the experimental data are from 
Refs.~\cite{d14C14p,d14C16p,d14C17p}.}
\end{figure*}

\begin{figure*}[!]
\begin{center}
\includegraphics[scale=\scl]{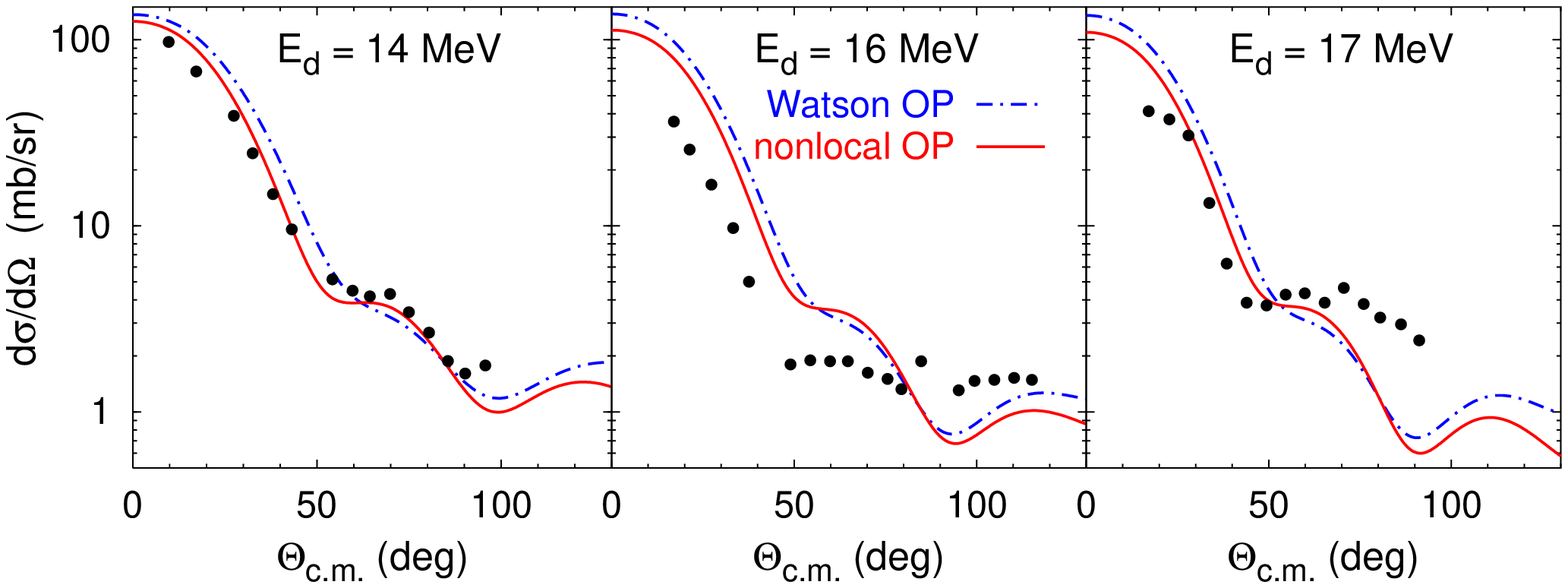}
\end{center}
\caption{\label{fig:dCpd}  (Color online)
Same as  Fig.~\ref{fig:dCps}, but for  transfer to the $\Cn$ 
excited state $5/2^+$.}
\end{figure*}

Results for $d+\C \to p+\Cn$ transfer at 
$E_d= 14$, 16, and 17 MeV are presented in Fig.~\ref{fig:dCps} for the $\Cn$ 
ground state $1/2^+$ and in Fig.~\ref{fig:dCpd} for the excited state $5/2^+$.
The corresponding proton lab energy in the time-reverse  reaction
$p+\Cn \to d+\C$ is $E_p = 12.0$ (11.2), 13.9 (13.1), and 14.8 (14.0) MeV
for the ground (excited) state, respectively.
The theoretical predictions vary smoothly with the energy while
 the data from Ref.~\cite{d14C16p} at 16 MeV are significantly lower
than other two sets from Refs.~\cite{d14C14p,d14C17p} at 14 and 17 MeV
as already pointed out in  Ref.~\cite{lee:07a}.
This raises serious doubts on the normalization of the 16-MeV data.
In all cases NLOP reproduces the shape of the data better than Watson OP,
especially for the transfer to the ground state $1/2^+$ in the region of
the first minimum and second maximum.
At 14 MeV and at 17 MeV up to c.m. scattering angle
$\Theta_{c.m.} \approx 60^{\circ}$ also
the quantitative description of the data by the NLOP is quite satisfactory.
The extracted neutron spectroscopic factor for the $\Cn$ ground state $1/2^+$ 
is close to 1, in agreement with the DWBA-type result of Ref.~\cite{lee:07a},
while for the  excited state $5/2^+$ it is slightly below 1.

In addition we performed the nearside-farside
decomposition of the cross section \cite{fuller:nf}. Although at small
angles where the cross section is largest the nearside usually dominates, the 
nonlocality effect may be equally important for both nearside and farside 
cross sections and therefore the observed nonlocality effect is a result of 
a complicated interference between nearside and farside.

The results of the present work together with those of Ref.~\cite{deltuva:09b}
indicate that the OP nonlocality effect on transfer observables
depends quite strongly on the reaction energy and internal orbital angular 
momentum of the $(An)$ nucleus but far less on its binding energy and mass.
The dependence on the internal orbital angular momentum is evident,
and the dependence on the reaction energy is seen most clearly
by comparing $d+\Be \to p+\Ben$ and $p+\Ben \to d+\Be$ results
at $E_d=12$ and 25 MeV and $E_p=35.3$ MeV given in  Figs.~\ref{fig:dBeps} and 
\ref{fig:pBe35d}. However, as shown in Figs.~\ref{fig:dBeps} 
and \ref{fig:dCps}, the OP nonlocality effect is very similar in
$(d,p)$ reactions at $E_d = 12$ and 14 MeV leading to the ground state $1/2^+$
of $\Ben$ and $\Cn$, respectively. Furthermore,
the OP nonlocality effect in the $p+\Ben \to d+\Be$ transfer at $E_p=35.3$ MeV
is very similar to the one shown in Figs.~5 and 7 of Ref.~\cite{deltuva:09b} 
for $(d,p)$ reactions at comparable c.m. energies leading to 
the excited state $1/2^+$ of $\A{13}{C}$ and  $\A{17}{O}$ nuclei
with the binding energies of 1.857 and 3.272 MeV, respectively.
There is also a mutual similarity between OP nonlocality effects
given in Figs.~5 and 6 of Ref.~\cite{deltuva:09b} for
 $(d,p)$  reactions around $E_d=30$ MeV leading to $5/2^+$ state of $\A{13}{C}$ 
and  $\A{17}{O}$ nuclei; the corresponding binding energies are 1.092 and 
4.143 MeV.

As in Ref.~\cite{deltuva:09b} we checked that the OP nonlocality effect observed
in transfer reactions is insensitive to small variations of  the parameter 
$w_N$ and therefore is not a consequence of only approximate on-shell
equivalence between NLOP and Watson OP. 
We also studied sensitivity of the results with respect to $np$ and $nA$ 
interactions.
We performed additional test calculations using local AV18 potential 
\cite{wiringa:95a} for $np$; compared to mildly nonlocal CD Bonn potential
the differences in the predictions were found to be entirely negligible.
Varying parameters $r_0$ and $a$ of the $nA$ potential within 
reasonable limits, $\pm 0.05$ fm, or, in the case of $n$-$\Be$, using
$L$-independent $V_{so}$, and adjusting the strengths 
$V_c$ and $V_{so}$ to reproduce the same binding energies, leads to
the differences in transfer results that are still considerably
smaller than the observed OP nonlocality effect.
Thus, the cross section in the considered  transfer reactions depends
 mostly on the $pA$ interaction.

\section{Summary} \label{sec:sum}

We performed calculations of $(d,p)$ and $(p,d)$ transfer reactions involving
one-neutron halo nuclei $\Ben$ and $\Cn$. 
Exact three-body scattering equations in
the AGS form were solved and the Coulomb interaction was included using
the method of screening and renormalization; well converged results were 
obtained. With respect to dynamics,
they are mostly sensitive to the proton-core interaction.
For the first time in reactions with halo nuclei a nonlocal optical 
potential was used and important nonlocality effects were found.
The OP nonlocality effect on transfer observables depends mostly on the 
reaction energy and internal orbital angular momentum of the halo nucleus 
and is rather insensitive to its binding energy and mass.
The NLOP accounts for the experimental transfer data better 
than the local Watson OP and provides a more consistent description
over a broader energy and angular range, especially for the $1/2^+$ 
initial/final state of the halo nucleus. 
The values for the neutron spectroscopic factor obtained using the NLOP 
agree quite well with estimates based on other approaches,
in contrast to those obtained using the nearly on-shell equivalent Watson OP.

\vspace{1mm}
\begin{acknowledgments}
The author thanks A.~C.~Fonseca for comments on the manuscript.
The work is supported by the Funda\c{c}\~{a}o para a Ci\^{e}ncia
e a Tecnologia (FCT) grant SFRH/BPD/34628/2007.
\end{acknowledgments}



\end{document}